\documentclass[3p,times,twocolumn]{elsarticle}

\usepackage{ecrc}

\volume{00}

\firstpage{1}

\journalname{Nuclear Physics B Proceedings Supplement}

\runauth{}

\jid{nuphbp}

\jnltitlelogo{Nuclear Physics B Proceedings Supplement}

\usepackage{amssymb}

\usepackage[figuresright]{rotating}

\newcommand{\MET}{E\llap{/\kern1.5pt}_T}

\begin{document}

\begin{frontmatter}



\dochead{}

\title{Disappearing charged tracks in association with displaced leptons\\from supersymmetry}


\author{Christoffer Petersson}

\address{Physique Th\'eorique et Math\'ematique, Universit\'e Libre de Bruxelles,
 C.P.~231, 1050 Brussels, Belgium\\
International Solvay Institutes, Brussels, Belgium\\
Department of Fundamental Physics, Chalmers University of Technology,
412 96 G\"oteborg, Sweden }

\begin{abstract}
In this note we discuss a characteristic collider signature of models of gauge mediated supersymmetry breaking in which the selectron and smuon are mass-degenerate co-NLSP. In these models, all processes involving superpartners give rise to two NLSP selectrons or smuons, each of which subsequently decays to a nearly massless LSP gravitino and an electron or a muon. In a large region of the parameter space, the NLSPs travel macroscopic distances, of the order $0.1{-}1000\,$mm, before they decay. Hence, these models give rise to collider signatures involving charged tracks that end at vertices, which are inside the detector volume but displaced with respect to the original collision point, from which an electron or a muon is emitted. In order to probe this class of models we propose a search for disappearing charged tracks in association with  displaced electrons or muons.  

\end{abstract}


\end{frontmatter}


\section{The selectron NLSP scenario}

While the wide range of LHC searches for new physics has placed strong bounds on colored superpartners, the electroweak superpartners remain less constrained, mainly due to the lower production cross-sections and difficult backgrounds. Moreover, most LHC searches assume that the decays of the superpartners are prompt. In this note we focus on pure electroweak production and point out a signature that involves superpartners with significant lifetimes, which give rise to charged tracks and displaced vertices. 

The signature we consider arises in a  relatively unexplored class of models based on gauge mediated supersymmetry breaking (GMSB) in the minimal supersymmetric standard model (MSSM). The non-standard feature of these models is that the mass-degenerate right-handed selectron and smuon, $\tilde{\ell}_R{=}\tilde{e}_R,\tilde{\mu}_R$, are the next-to-lightest superpartners (NLSP). Note that in more well-studied versions of GMSB, the NLSP is instead either the lightest stau mass eigenstate  
or the lightest neutralino 
\cite{Giudice:1998bp}. The lightest superpartner (LSP) is, as in any GMSB model, a nearly massless gravitino, $\tilde{G}$. For brevity, we refer to this exotic GMSB spectrum as the selectron NLSP scenario.\footnote{Note also the distinction from the so-called slepton co-NLSP scenario, which refers to the spectrum where the right-handed selectron, smuon and stau are all nearly mass-degenerate \cite{Ruderman:2010kj}. Here, instead, in the selectron NLSP scenario, the stau is parametrically heavier than the selectron and smuon.}  

A complete characterisation of GMSB models realizing the selectron NLSP scenario is provided in \cite{Calibbi:2014pza}. This spectrum can be achieved in explicit weakly coupled messenger models that allow for direct couplings between the messenger fields and the MSSM Higgs sector \cite{GMSBwA}. Moreover, these models have the attractive features of accommodating large $A$-terms, relatively light stops and a Higgs mass of 126 GeV without severe fine-tuning. The desired splitting between the first two slepton generations and the third generation  is induced by the usual flavor textures of the Standard Model Yukawa couplings. Hence, the selectron NLSP scenario is achieved without introducing any new source of flavor misalignment. 

We assume R-parity conservation and therefore, the only decay mode for the selectron/smuon NLSP is the 2-body decay to the gravitino and an electron/muon, with the decay width given by \cite{Giudice:1998bp},
\begin{equation}
\label{NLSP}
\Gamma ( \tilde{\ell}_R \to \ell \tilde G )=\frac{m_{\tilde{\ell}_R}^5}{48 \pi \,m_{3/2}^2 M_{P}^2}
\end{equation}
 where $\ell{=}e,\mu$, $M_P{=}2.4{\cdot}10^{18}\,$GeV and $m_{3/2}$ is the gravitino mass. 
 
\begin{figure}
\begin{center}
\includegraphics[width=3in]{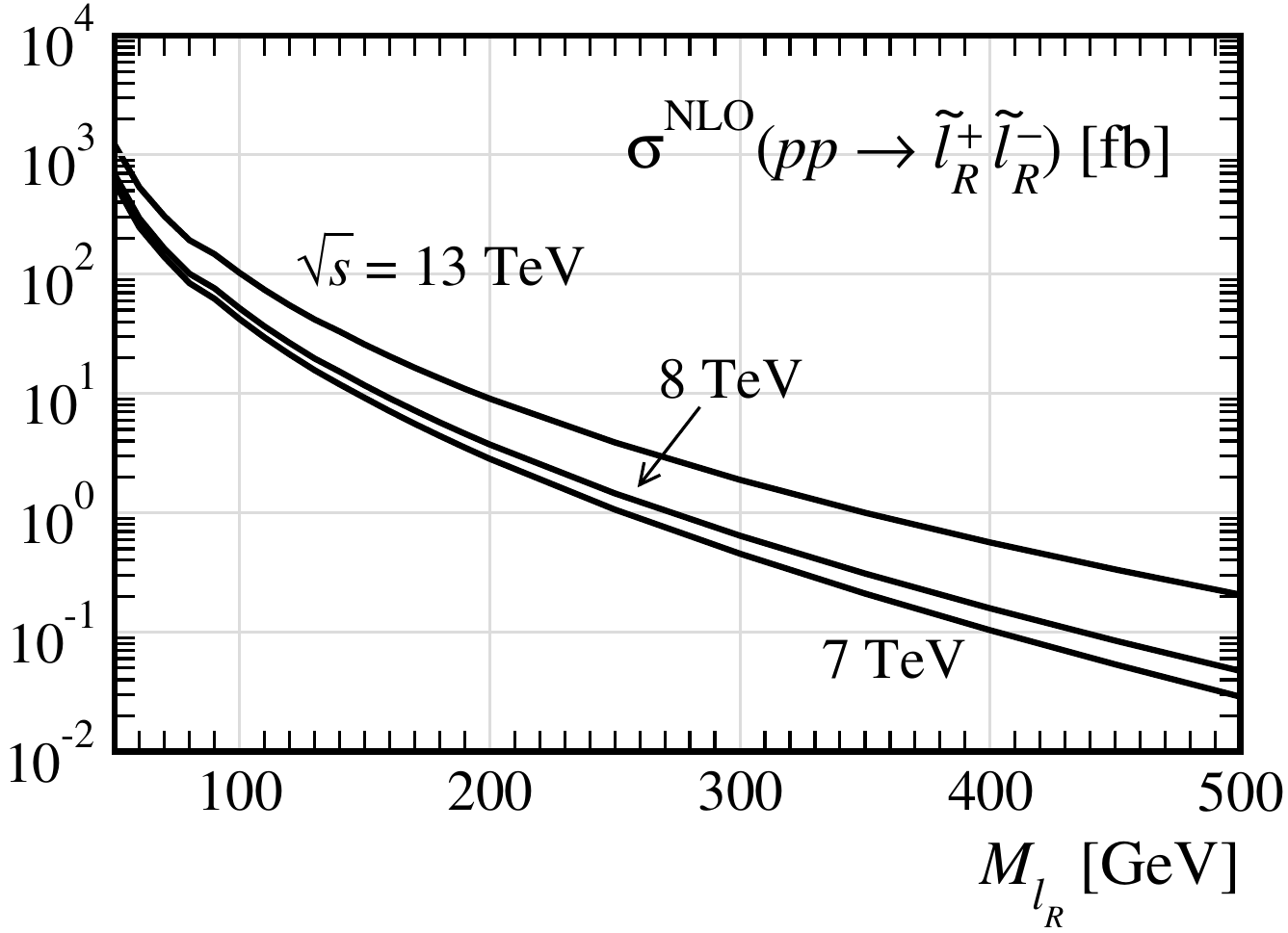}
\caption{Pair production cross section at the LHC for a single flavor right-handed sleptons.}
\label{slepxsec}
\end{center}
\end{figure}

 At the LHC, the selectron/smuon NLSP can be pair produced via Drell-Yan production, $pp {\to} Z/\gamma {\to} \tilde{\ell}_R\tilde{\ell}_R$. The cross-section at $\sqrt{s}\,{=}\,7,$ 8 and 13\,TeV, at next to leading order in QCD, is given in Figure \ref{slepxsec}, as computed by MadGolem \cite{Binoth:2011xi}.    

 The superpartner above the selectron/smuon NLSP in the spectrum is typically the lightest stau.  However, the masses of the stau and the other superpartners are model-dependent and, in fact, all the remaining superpartners (including the lightest stau) can be parametrically heavier than the selectron/smuon NLSP \cite{Calibbi:2014pza}. Therefore, in order to reduce the number of free parameters down to only the most important ones, we take the simplified model approach and take all the superpartners above the selectron/smuon NLSP to be effectively decoupled. In this case, there are only two parameters, the NLSP mass of the degenerate selectron/smuon, $m_{\tilde{\ell}_R}$, and the gravitino mass, $m_{3/2}$. These two parameters determine the NLSP production cross-section and lifetime. 

\section{Collider signature and existing LHC searches}    

As can be seen from Eq.~(\ref{NLSP}), depending on the masses of the selectron/smuon NLSP and the gravitino, the NLSP decay can be prompt or long-lived on collider time scales, or, in the intermediate scenario, it can give rise to a displaced vertex.\footnote{For a discussion concerning displaced NLSP decays and LHC searches in the context of the pMSSM, see \cite{Cahill-Rowley:2013yla}.} It turns out that, in a large region of the parameter space of the models that realize the selectron NLSP scenario, the NLSP and the gravitino masses are in the ranges $100{-}500$\,GeV and $10{-}10^4$\,eV, respectively, implying that the average distance the selectron/smuon NLSP travels before it decays is in the range $0.1{-}1000\,$mm \cite{Calibbi:2014pza}.  Hence, the collider signature involves two charged tracks, each of which ends at a vertex (that is displaced with respect to the original collision point) from which an electron or a muon originates. Moreover, from each of the displaced vertices an invisible gravitino is emitted, giving rise missing transverse energy $(\MET)$.  

Before we further discuss this exotic signature let us briefly discuss the two other (more familiar) lifetime cases, i.e.~when the NLSP decay is either prompt or long-lived. If the NLSP decay had been prompt\footnote{A simplified model where the selectron and smuon are co-NLSP and decay promptly was employed by CMS in \cite{Chatrchyan:2014aea}, see also \cite{D'Hondt:2013ula,Petersson:2014faa} for further discussions.}, the ATLAS search \cite{Aad:2014vma} in the final state \mbox{$\ell^+\ell^- {+}\MET$} would have constrained the right-handed selectron/smuon NLSP mass to be above 245\,GeV.\footnote{Note that the NLSP decay is prompt in the case where the gravitino mass is small, or, equivalently, where the supersymmetry breaking scale is low. See \cite {lowscale} for some recent discussions concerning low scale supersymmetry breaking.} If instead the selectron/smuon NLSP had been long-lived, i.e.~if the NLSP decay would take place outside the detector, then the  NLSPs would appear as heavy muons, giving rise to two charged tracks stretching through the entire detector. Both ATLAS and CMS have  searched for such long-lived charged particles  \cite{longlived} and their results would have constrained the selectron/smuon NLSP mass to be above around 400\,GeV \cite{Calibbi:2014pza}.

 Let us now turn to the intermediate case we are interested in, i.e.~where the NLSP decays inside the detector. To the best of our knowledge, there is no existing LHC search for this case. However, as we will review below, there are several searches involving displaced vertices. We will propose that, by combining or extending these existing searches, it would be possible to design a search optimized for probing the selectron NLSP scenario.   
 
First of all, if the displaced electrons or muons, which arise from the NLSP decays in our signal process, would not be reconstructed, then the NLSPs would give rise to two charged tracks that ``disappear'' at some point in the detector. Such disappearing tracks have been searched for by ATLAS  \cite{Aad:2013yna} but, unfortunately, this analysis involves a jet requirement of at least one jet with $p_{T}{>}90\,$GeV. The jet requirement implies that this analysis has negligible sensitivity to the selectron NLSP scenario, in which the only relevant production mode is pure electroweak production and the hadronic activity is small. 


If instead the charged tracks in our signal process would not be reconstructed, then the NLSPs would give rise to two displaced electrons or muons. ATLAS has searched for a heavy neutral particle that gives rise to at least one displaced vertex from which a muon and multiple tracks originate \cite{TheATLAScollaboration:2013yia}. In a signal region where they require a muon with $p_T{>}55\,$GeV and at least four other tracks, originating from the same displaced vertex, they set a 95\% CL bound on the visible cross section $\sigma\times\mathcal{A}\times \epsilon \,{<} \,0.14$\,fb, where $\sigma$ is the production cross section for the signal process, $\mathcal{A}$ is the detector acceptance and $\epsilon$ is the reconstruction efficiency. Of course, there are two key differences between this process and our signal process. First of all, the intermediate particle that travels a macroscopic distance before it decays is neutral instead of charged. Secondly, our signal process involves the selectron/smuon NLSP decay to an electron/muon and an invisible gravitino. Therefore, there are very few tracks that originate from the displaced vertex and the $\mathcal{A}\times \epsilon$ for our signal process is vanishingly small. This, together with the fact that our process is characterised by a small (electroweak) production cross section, implies that this search is not expected to place any meaningful bound on the selectron NLSP scenario. 

Let us also mention the CMS search for events containing exactly one electron and one muon, originating from two separate displaced vertices \cite{Khachatryan:2014mea}. This process should be contrasted with our signal process, which arise from the production of a pair of selectrons or smuons, giving rise to two displaced leptons of the {\it same flavor}, i.e.~either two electrons or two muons, which originate from two separate displaced vertices.  Therefore we do not expect this search to constrain our signal process. 

From this discussion it is clear that an optimized search for the selectron NLSP scenario could be obtained by combining a search for disappearing charged tracks with a search for displaced leptons, and requiring the tracks and the leptons to meet at the displaced vertices.     
 
 \section{Conclusion}
  
In this note we have discussed a characteristic signature of a non-standard class of GMSB models in which the selectron and smuon are mass-degenerate NLSP. The selectron/smuon NLSP has a significant lifetime and travels a macroscopic distance before it decays inside the detector to an electron/muon and an invisible gravitino. This class of models can be probed by a search for disappearing charged tracks and associated displaced vertices from which an electron or a muon originates.

\section*{Acknowledgments}

I would like to thank L.\,Calibbi, A.\,Mariotti and D.\,Redigolo for the collaboration on the project I discussed here. 
This work is supported by the Swedish Research Council (VR) under the contract 637-2013-475, by IISN-Belgium (conventions 4.4511.06, 4.4505.86 and 4.4514.08) and by the ``Communaut\'e Fran\c{c}aise de Belgique" through the ARC program and by a ``Mandat d'Impulsion Scientifique" of the F.R.S.-FNRS. Finally I would like to thank the organizers of the conference ``ICHEP 2014" in Valencia for their efforts in organizing this nice and interesting conference.







\end{document}